# Binding of native DNA to MoS$_2$ nanoflakes: the role of defects and edge atoms of MoS$_2$ nanostructures in their biofunctionalization

Alexander Glamazda[a,b,*], Evgeniya Usenko[a], Anastasiia Svidzerska[a], Vladimir Valeev[a], Igor Voloshin[a], Anna Laguta[b], Sergey Petrushenko[b,c], Stepan Stepanian[a], Ludwik Adamowicz[d], and Victor Karachevtsev[a]

[a] B. Verkin Institute for Low Temperature Physics and Engineering of the National Academy of Sciences of Ukraine, 47 Nauky Ave., Kharkiv 61103, Ukraine

[b] V.N. Karazin Kharkiv National University, Kharkiv 61022, Ukraine

[c] Technical University of Liberec, Liberec 461 17, Czech Republic

[d] Department of Chemistry and Biochemistry, University of Arizona, Tucson, AZ 85721, USA

*Corresponding author: glamazda@ilt.kharkov.ua

**Abstract**

In this work, the binding of native DNA to $MoS_2$ nanoflakes (FLs) was studied by using UV-visible absorption spectroscopy, thermal denaturation method, transmission electron microscopy (TEM), temperature-dependent dynamic light scattering (DLS), and the DFT computational-chemistry method. Analysis of the experimental data: TEM images and thermal denaturation measurements showed the binding of the biopolymer with $MoS_2$ FLs. An increase in the melting temperature of DNA and a decrease in the hyperchromic coefficient at binding with $MoS_2$ FLs indicates the formation of the DNA:$MoS_2$ FL nanoassemblies due primarily to the covalent interaction of the oxygen atoms of the phosphate groups of DNA with the $MoS_2$ FLs. Possible complexes of a nucleotide fragment (ribose-phosphate group) with $MoS_2$ nanolayer are considered and calculated employing the DFT method. Different structures of these complexes are optimized and the interaction energies between components are determined. Special attention in calculations is focused on the binding of this nucleotide fragment with Mo atoms located at the edge of the $MoS_2$ nanolayer and with point structural defects of the $MoS_2$ surface containing the S vacancy. Based on this calculation and experimental observation, a mechanism of binding of native DNA to $MoS_2$ FLs has been proposed, in which their conjugation begins with point contacts of DNA phosphate groups with Mo atoms (at the edge or/and in defects) through the formation of a strong coordination bond. The results indicate the critical role of defects and edge atoms of $MoS_2$ FLs in their biofunctionalization.

# 1. Introduction

Recently, DNA sensors have been attracting much attention due to their high sensitivity, selectivity, and low cost [1]. Detection of DNA targets has great scientific and practical importance and is vital in medical diagnostics, cancer research, forensics, environmental monitoring, etc. [2-5]. Currently, DNA sensors are being developed in combination with various nanomaterials, which for the most part have outstanding physicochemical properties that are not inherent to bulk materials. Among such materials, transition metal dichalcogenides (TMDs) possess a high surface-to-volume ratio, tunable bandgap, strong spin-orbital coupling, and high optical and thermal conversion efficiency. These properties open a wide range of applications of these materials as sensing platforms [6-8]. The TMD-based sensors have emerged as a promising alternative in the search for more efficient, fast, and reliable methods for detecting biochemical components in biomedical applications, environmental monitoring, and food safety.

Among all known TMDs, molybdenum disulfide ($MoS_2$) has attracted particular interest due to its exceptional optoelectronic coupling and catalytic, energetic, and biological properties [6,7]. In particular, $MoS_2$ layers can act as platforms for immobilized biomolecules by exploiting $MoS_2$ or dye fluorescence quenching to detect proteinsand nucleic acids [9-12]. A single $MoS_2$ layer consists of a molybdenum layer sandwiched between two sulfur layers [13]. $MoS_2$ can also be controllably synthesized with desired morphologies such as flakes (FLs) consisting of a few layers, quantum dots, radar-like particles, flower-like particles, and nanotubes. These structures can be used in various applications [9,10]. Interestingly, bulk $MoS_2$ has an indirect bandgap (1.2 eV), while exfoliated $MoS_2$ exhibits a direct bandgap (1.8 eV) and, thus, exhibits absorption and photoluminescent properties [11,12]. These unique structural, optical, and physicochemical characteristics allow the use of $MoS_2$ in the fields of catalysis, energy, electronics, medicine, and biophysics [14-17]. The induced defects on the $MoS_2$ surface play an important role in tuning electronic and optical properties, improving photocatalytic activity, etc. Thereby it extends the scope of the application of the modified $MoS_2$[18,19].

It was experimentally and theoretically demonstrated that $MoS_2$ nanopores are capable of distinctly detecting single nucleotides. In general, the $MoS_2$ nanopore systems demonstrate high selectivity towards individual molecules. However, they have not yet found any significant application in clinical medicine due to the complexity involved in manipulating them [20,21]. Nevertheless, there have been studies devoted to creating new electrochemical biosensors based on water-soluble $MoS_2$ materials for the detection of DNA [22,23]. In Refs. [24-28], data on the spectral and current-voltage characteristics of MoS$_2$ are provided, as well as of its complexes with various species. To employ MoS$_2$ nanosheet as a sensing platform for a nucleic acid sequence detecting method a comprehensive study of the interaction between DNA and MoS$_2$ should be provided. The recently performed experimental [29-32] and theoretical [33-39] studies showed that the binding of DNA to the MoS$_2$ FL surface occurs through van der Waals forces (see also Review [40, 41]). The hydrophobic interactions and hydrogen bonds play only a small role in DNA adsorption onto the MoS$_2$ FL surface [29]. The basic strategy discussed in the abovementioned experimental works is based on the adsorption of short dye-labeled single-stranded oligonucleotide (ssDNA) on a single-layer or a few layered MoS$_2$ as a probe. The DNA adsorption in these experiments is controlled by the quench of the dye fluorescence. In the case of hybridization with complementary target oligonucleotide, the duplex formed can detach from the surface of MoS$_2$ as the duplex binding to the surface is weaker versus ssDNA [32].

The interaction between single nucleobase/nucleotide/polynucleotides and MoS$_2$ was investigated through DFT calculations and molecular dynamics (MD) simulation [31-37]. From the performed calculation it follows [33-36, 38], that guanine and adenine generally have larger binding energies than cytosine and thymine. This trend (as derived from DFT calculations [37]) is maintained for nucleotides adsorbed to MoS$_2$ (G > A > C > T). As it was recently shown by molecular dynamics simulation, a short double-stranded DNA (dsDNA) can be adsorbed on the MoS$_2$ surface effectively only when dsDNA adopts vertical orientation and can interact only through terminated nucleobases stacked with the nanosheet's surface and stabilized by weak van der Waals forces [42]. In another MD simulation [43], the adsorption of short-length ssDNA (ten nucleobases repeat) on a single-layer MoS$_2$ surface in aqueous solution is considered. It was shown that the oligonucleotides adsorption on MoS$_2$ via the π–π stacking interactions inducing conformational changes in the ssDNA. Heterostructures, namely (AC)$_5$ and (GT)$_5$, demonstrate competition between π-stacking interactions: nucleobase–nucleobase (intramolecular) and nucleobase–MoS$_2$ (intermolecular). The heterooligonucleotides adopt relatively compact duplex conformation on the MoS$_2$ surface although with significant structural transformation. It was noted that the self-stacking between the nucleobases reduces van der Waals interaction with the MoS$_2$ surface [43].

Recently, the stacked and covalently bound nanostructures formed by nucleobases with MoS$_2$ monolayer fragments have been calculated, and the structures and interaction energies were determined [36]. It was shown that the covalent interaction of bases with edges of the MoS$_2$ monolayer leads to a significant increase in the binding energy, which is accompanied by a spectral transformation of the vibrational spectra [38, 39]. It should be noted that, despite of abovementioned investigations, more research in this area is needed, especially with regard to the binding of long and native DNA to $MoS_2$ nanostructures or there is also no information on the effect of the temperature on the interaction between DNA and $MoS_2$ when dsDNA is denaturated. It is now known that one of the stages of DNA replication in a living cell is the division of a double strand of DNA into two single strands due to the breaking of the hydrogen bonds between the paired bases [44]. A special enzyme called helicase regulates this process. In laboratory conditions, the double-strand separation can be achieved by increasing the temperature. Note that thermal denaturation helps to study the stability of DNA under different conditions and is an analog of the action of the helicase enzyme in DNA unwinding [45].

In the present work, we study whether native DNA would bind to $MoS_2$ FLs at relatively low ionic force ($10^{-3}$ $Na^+$, pH7) using experimental and theoretical methods. Based on analysis of the TEM image which showed the adsorption of DNA on $MoS_2$ FLs and the observation of an increase in the thermal stability of DNA and a decrease in the hyperchromic coefficient at binding with $MoS_2$ FLs in the thermal denaturation measurements we concluded about the binding of the biopolymer with $MoS_2$ nanoflakes at these conditions. It was observed, first of all, that the melting curve of DNA in the nanoassemblies has S-like form that indicates keeping the duplex structure of DNA at the conjugation. Observed changes in these DNA characteristics (an increase in $T_m$ and a decrease of $h$) were explained by the electrostatic interaction of the negatively charged oxygen atoms of the phosphate groups of DNA with the $MoS_2$ FLs. Consideration of possible complexes of a nucleotide fragment (ribose-phosphate group) with $MoS_2$ nanolayer employing the DFT method showed the forming of the coordination bonds of this nucleotide fragment with Mo atoms located at the edge of the $MoS_2$ nanolayer and with point structural defects of $MoS_2$ surface containing the S vacancy. This observation allows us to propose a mechanism of binding of native DNA to $MoS_2$ FLs, in which their conjugation begins with point contacts of DNA phosphate groups with Mo atoms (at the edge or/and in defects) through the formation of the coordination bond. The results indicate a huge impact of defects and edge atoms of $MoS_2$ FLs on their biofunctionalization.

## 2. Materials and methods

*2.1. Materials*

The MoS$_2$ powder purchased from Sigma-Aldrich (Product Number: 234842) is used in the present work. The exfoliation of the MoS$_2$ bulk material is performed in accordance with reports [46-48]. It involves dispersing 15 mg of MoS$_2$ powder in an ammonia-water solution (5ml) with pH of 9.80. The MoS$_2$:NH$_3$ aqueous suspension is ultrasonicated with $v$=22 kHz for 200 min at room temperature. Next, the suspension is centrifuged for 30 minutes at the speed of 2000 rpm. The applied "soft" method of the exfoliation of the MoS$_2$ materials does not damage the basal planes [47]. The ionic environment and the edge-attached groups improve the dispersion of the exfoliated MoS$_2$ material. The supernatant is collected and further used in the study. The MoS$_2$ concentration is ≈20 μg/ml. The process for preparing the ammoniated MoS$_2$ material with DNA involves the following steps: 1) the calf-thymus DNA with a molecular weight of 1.2×10$^6$ Da (Serva, Germany) is suspended in a buffer solution containing 10$^{-3}$ M sodium cacodylate (CH$_3$)$_2$AsO$_2$Na•3H$_2$O (Serva, Germany) at pH7; 2) the necessary quantity of MoS$_2$ is introduced into the DNA buffer suspension. The concentration of MoS$_2$ and polynucleotide phosphorus can be determined with an error margin of no more than 0.5%.

The phosphorous concentration of DNA, denoted as [P], is determined to be (6.61±0.2)×10$^{-5}$ M. This value was obtained based on the DNA molar extinction coefficient ($\varepsilon_m$=6600 M$^{-1}$×cm$^{-1}$) taken at the frequency $v_m$=38500 cm$^{-1}$ [49].

*2.2. Transmission electron microscopy (TEM)*

We have studied morphology of MoS$_2$ FLs and DNA:MoS$_2$ FL nanoassemblies using a SELMI EM-125 (Sumy, Ukraine). The prepared samples were condensed onto freshly cleaved KCl single crystals with an amorphous carbon sublayer. A 3-μL of sample was dropped on the freshly cleaved KCl single crystals with an amorphous carbon sublayer and dried for analysis.

*2.3. Dynamic light scattering*

The size distributions of MoS$_2$FLs and DNA:MoS$_2$ FL nanoassemblies are determined via dynamic light scattering (DLS) using Zetasizer Nano ZS Colloid and Polymer Science (Red badge) ZEN 3600 Malvern Instruments apparatus at temperatures ranging from 25 to 90 °C. The DLS measurements are carried out at 632.8 nm (output power of 4 mW) with a He-Ne laser at a 175° scattering angle. The temperature in the cuvette holder is controlled automatically and increased in steps of 2 °C. The duration of the isotherm is 30 s (except for the measurements at 25 °C, which is the first point, the remaining measurements are carried out with an equilibrium time of 120 s). Size measurements are conducted using a glass cuvette, and zeta-potential measurements are carried out using disposable folded capillary cells. The methodology for DLS measurements is given in Ref. 50

*2.4. Differential UV and visible spectroscopy*

Differential absorption UV (DUV) spectroscopy is employed for detecting subtle alterations in the electronic/structural configuration of DNA caused by its surroundings. Specord UV-vis spectrophotometer (Carl Zeiss, Jena, Germany) is used to record DUV spectra of the DNA:MoS$_2$ FL suspensions, employing the four-cuvette experiment. The 1$^{st}$ and 3$^{rd}$ cuvettes are used for the DNA buffer suspension and buffer solution of samples, respectively. The 2$^{nd}$ and 4$^{th}$ cuvettes contain DNA and buffer solution identical to the 1$^{st}$ and 3$^{rd}$, respectively. The effect of MoS$_2$ on structural changes in DNA was assessed by injecting an equal amount of MoS$_2$ FLs into the 1$^{st}$ and 4$^{th}$ cuvettes. The light absorption changes, represented by $\Delta\varepsilon_h(v)$, are normalized to the concentration of polynucleotide phosphates using the equation $\Delta\varepsilon_h(v)=\Delta A(v)/[P]$, where $\Delta A(v)$ is the relative change in the optical density. The sample degradation was not observed during the experiment.

*2.5. Thermal denaturation*

Thermal measurement analysis helps in identifying the DNA structural stability and various thermodynamic binding parameters of the ligands to DNA, including the melting temperature ($T_m$), width of the helix-coil transition interval, etc.[51]. The stability of the biopolymer structure is characterized by $T_m$. When the temperature is raised, the native DNA is disrupted, causing the biopolymer to create loops due to the breaking of hydrogen bonds between complementary base pairs. When the temperature approaches $T_m$, half of the DNA exists in a helical structure, while the remaining half exists as individual strands. The increase in $T_m$ is caused by the structural stabilization of the biopolymer. We have measured the optical density of the DNA suspensions at a fixed energy of 38500 cm$^{-1}$ in the dependence on temperature. This dependence describes the melting curve with an S-shaped anomaly at $T_m$.

In the context of our studies, we employ the lab software to record the melting curves, which represent the temperature-dependent changes in the hyperchromicity coefficient: $h(T)=[\Delta A(T)/A_{To}]v_m$. Here, $\Delta A(T)$ denotes the alteration in the optical density of the DNA suspension when heated, and $A_{To}$ represents the optical density at $T=To$. Hence, $h(T)$ serves as a quantitative measure of hyperchromism.

*2.6. Computational details*

The geometries of the MoS$_2$-dMP (deoxyribose 1-methyl-5-monophosphate) complexes are calculated at the DFT level of theory. The dMP molecule mimics the sugar-phosphate backbone of DNA and is an analog of nucleotides in which the nucleic bases are replaced by methyl groups. It is used to estimate the interaction energy and the preferred binding sites of a double-stranded DNA with a MoS$_2$ layer. Nucleic bases in a double-stranded DNA are less accessible to form intermolecular interactions as compared to a single-stranded DNA or a single nucleotide. Therefore, the use of the model molecule (dMP) is only usefulin determining the parameters of the interaction of MoS$_2$ with the side chain of a double-stranded DNA. The DFT calculations are performed with the M06-2X density functional [52]. This functional enables to account for the dispersion interactions without using empirical corrections. The counterpoise corrections are calculated using the standard counterpoise correction procedure [53]. For the dMP atoms, we use the standard 6-31++G(d,p) basis set. For the Mo and S atoms, we use the LanL2DZ and LanL2DZdp ECP basis sets, respectively [54-57].

A fragment of the pristine MoS$_2$ 2D layered surface which included 27 molybdenum atoms and 54 sulfur atoms is used in the calculations. The calculations are carried out for two stacked complexes designated as MoS$_2$-dMP S and for two covalently bonded complexes designated as MoS$_2$-dMP B. The initial structures of these complexes differ in the location of dMP on the fragment of the MoS$_2$layer. In the stacked complexes, the dMP molecule is completely located on the MoS$_2$ layer surface formed by the sulfur atoms. In the MoS$_2$-dMP B complexes, the dMP molecule is located at the edge of the MoS$_2$ fragment, forming covalent bonds with accessible molybdenum atoms. In addition to the dMP complexes with a pristine MoS$_2$ layer fragment, we also study complexes involving MoS$_2$ fragments with point structural defects. These types of defects are often observed in 2D MoS$_2$ [58, 59]. Two types of defects in the MoS$_2$ layer are considered in the calculations: the Mo$_S$ defect (one Mo atom replacing the S site) and the V$_{2S}$ defect (two S atoms missing). The structures of the defects are shown in Scheme 1. The structures of all considered complexes are fully optimized. This is followed by harmonic frequency calculations to confirm that the structures found correspond to true minima on the potential energy surfaces of the complexes. The harmonic frequencies are also used to calculate the ZPVE corrections to the interaction energies. Similar calculations are also carried out for all complexes embedded in aqueous environment. The Polarizable Continuum Model (PCM) approach (integral equation formalism model (IEFPCM) [60]) is used in the calculations. Since the BSSE corrections cannot be calculated in combination with the PCM approach, BSSE corrections are calculated without taking into account the aqueous environment for the geometries fully optimized using the M06-2X/PCM method.

The Gaussian 16 software package is used for performing calculations in the present manuscript [61].

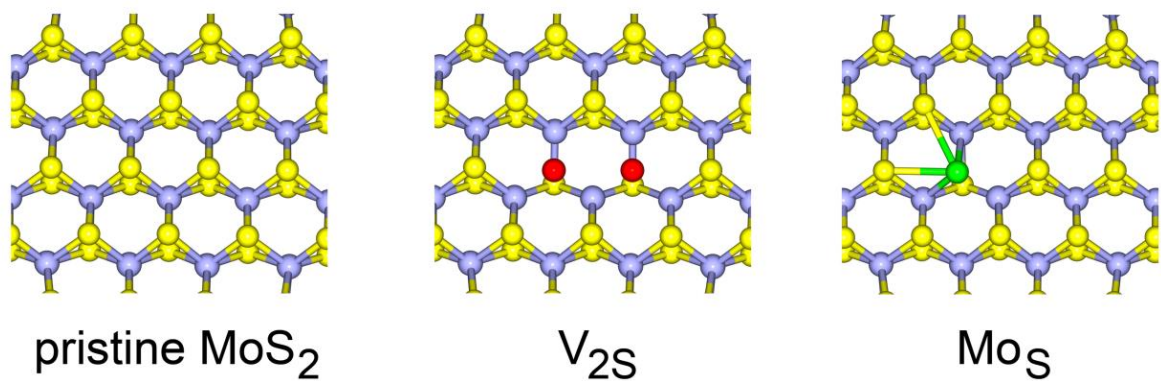

**Scheme 1.** Schematic representation of the pristine MoS$_2$, as well as MoS$_2$ with V$_{2S}$ and Mo$_S$ defects. Removed sulfur atoms (V$_{2S}$) are shown in red. The substituting molybdenum atom (Mo$_S$) is shown in green.

## 3. Results and discussion

*3.1. TEM analysis*

The morphology of the prepared $MoS_2$ nanomaterial is studied with TEM. The performed analysis of TEM image in Fig. 1a reveals well-defined 2D multilayer FLs. A noticeable contrast between one FL and a few flaked aggregates can be observed. That is in accordance with the recently published TEM data concerning disordered $MoS_2$ FLs [62].

The morphology of the DNA:$MoS_2$ FL nanocomposites is also analyzed with TEM. As shown in Fig. 1b, the DNA:$MoS_2$ FL nanocomposites exhibit agglomerated functionalized nanostructures. The biopolymer cover of $MoS_2$ FLs blurs their outlines. The binding of DNA with $MoS_2$ FLs is further confirmed through spectroscopic studies.

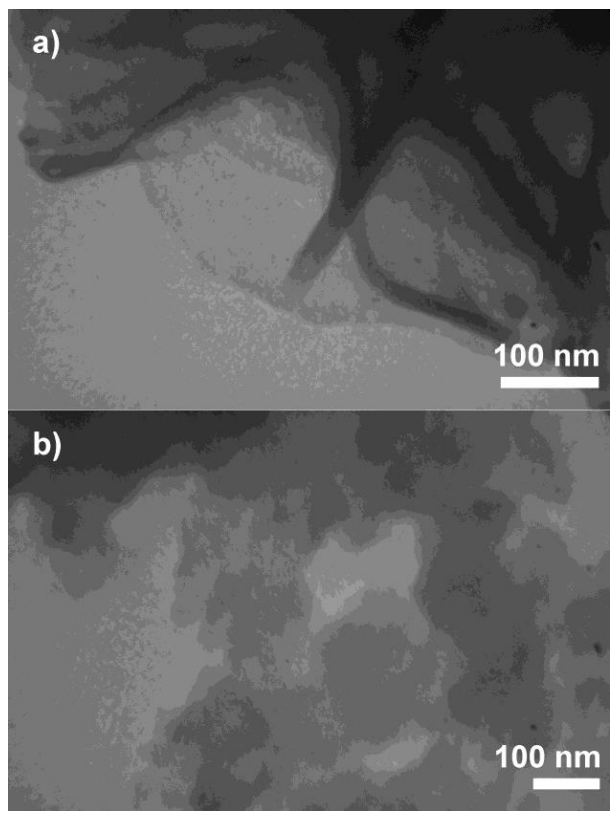

**Fig. 1.** TEM images of (a) $MoS_2$ FLs and (b) DNA:$MoS_2$ FL nanoassemblies.

*3.2. Characterization of MoS$_2$ FLs by absorption spectroscopy*

Figure 2 shows the UV–vis optical absorption spectra of the MoS$_2$ FL suspension with different concentrations. The spectra show that the absorption intensity increases as the MoS$_2$ FL concentration increases, indicating that there are no the aggregation processes with the MoS$_2$ FLs concentration. The observed absorption peaks at about 615 (B band) and 675 (A band) nm originate from direct excitonic transitions from the K point of the Brillouin zone in MoS$_2$ FLs. Two broad bands at about 400 (C band) and 450 (D band) nm correspond to direct exciton transitions at the M point in the high-density band-structure region. The absorption peak at about 260 nm can be assigned to excitonic features of the MoS$_2$ quantum dots (MoS$_2$ QDs) [63]. Thereby, the above results provide an indication that the preparation approach of MoS$_2$ FLs used in this work works. The observed spectral peculiarities are attributed to the well-separated MoS$_2$ layers and are not observed in the absorption spectra of the MoS$_2$ bulk material.

The average lateral size of MoS$_2$ FLs can be calculated by using the following equation [64]:

$$L(\mu m) = \frac{3.5 Ext_B / Ext_C - 0.14}{11.5 - Ext_B / Ext_C} \quad (1)$$

where $L$ is the average lateral size of MoS$_2$ FLs, $Ext_B$ is the extinction value of exciton B and $Ext_C$ is the extinction value of exciton C in the UV–vis spectra. The number of the MoS$_2$ layers per FL ($N$) can be measured by analyzing the following equation:

$$N = 2.3 \times 10^{36} \exp\left(-\frac{54888}{\lambda_A}\right) \quad (2)$$

where $\lambda_A$ – peak position of A band in nm. An analysis of our sample gives $L \approx 125$ nm and $N \approx 12$. Thus, the suspension contains MoS$_2$ FLs with this determined average size and MoS$_2$ QDs.

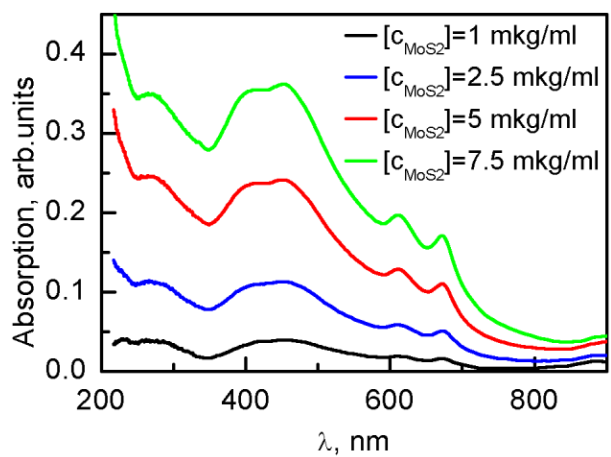

**Fig. 2.** Absorption spectra of MoS$_2$ FLs obtained at different concentrations.

*3.3. Thermal denaturation*

The typical temperature dependence of the hyperchromic coefficient (*h*) of DNA has the usual S-like shape: heating induces a helix-coil structural transition, which is reflected in the appearance of hyperchromism (*h* > 0) (Fig. 3, curve 1). An analysis of the DNA melting curve allows us to determine the thermodynamic parameters of the biopolymer: $T_m$=51.5 °C and $h_{m0}$=0.37 (where $h_{m0}$ - the value of the hyperchromic coefficient in the absence of MoS$_2$ FLs). However, the addition of MoS$_2$ FLs to the DNA buffer suspension does not cause any dramatic changes in the shape of the biopolymer melting curve. Each of the DNA melting curves presented in Fig. 3 obtained in the absence and the presence of MoS$_2$ FLs can be separated into several linear segments. The a-b segments are typical for double-strand melting regions. Their appearance may be explained by the fluctuation melting of the double helix that starts from the ends [65]. The b-c segment corresponds to splitting of the double-stranded DNA into two single strands (helix-coil transition). This transition ends at point C. Finally, the melting of single-stranded DNA fragments formed as a result of the splitting of the double-stranded DNA occurs (the c-d segment). Note that the DNA melting curve maintains the S-shape for the biopolymer in the nanoassembly indicating that the duplex structure of the DNA is preserved. This means that conjugation of DNA with MoS$_2$ FLs occurs mainly due to point contacts, which is not accompanied by duplex unwinding.

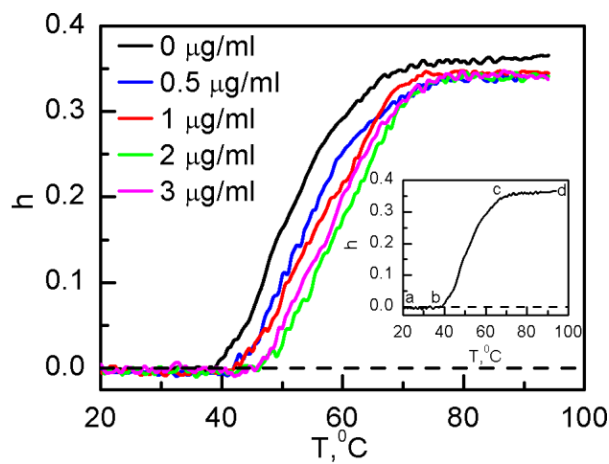

**Fig. 3.** Temperature dependence of the hyperchromic coefficient (*h*) of DNA without and with MoS$_2$ FLs. The inset shows the *h*(*T*) dependence of DNA without MoS$_2$ FLs. There are selected segments marked with letters (see the text).

The analysis of the melting curves shown in Fig. 3 allows us to obtain the concentration dependences of the melting temperature (Fig. 4a) and the hyperchromic coefficient (Fig. 4b) for the DNA:MoS$_2$ FL suspension. According to Fig. 4, we conclude that the injection of [$c_{MoS2}$]=2 µg/ml to the DNA suspension leads to an increase in the DNA melting temperature by 8.2 $^0$C. Note, that further addition of [$c_{MoS2}$]=3 µg/ml leads to a slight decrease in $T_m$ by 1.3 $^0$C. The increase in $T_m$ with MoS$_2$FL concentration can be explained by the interaction of positively charged atoms of MoS$_2$ FL with the negatively charged phosphate group of DNA. In this case, the repulsion between the phosphate groups located at the two opposite DNA strands weakens and DNA stability increases, as observed in the experiment. Such atoms of MoS$_2$ FL can be Mo atoms located at the nanoflake edge or Mo atoms exposed on the MoS$_2$ surface as a result of the structural defects of the surface containing an S vacancy.

As follows from Fig. 4b, the value of *h* is 0.37 for pure DNA, and with the addition of [$c_{MoS2}$]=0.5 µg/ml *h* decreases by 8% and becomes equal to 0.34. The value of *h* is maintained up to a concentration of 3 µg/ml. This means that, when molybdenum disulfide is added to the DNA suspension, some unwound single-stranded DNA regions may appear, but overall the double-stranded conformation is maintained when DNA interacts with MoS$_2$ FLs. The binding of DNA with MoS$_2$ FL can occur through a point contact at room temperature.

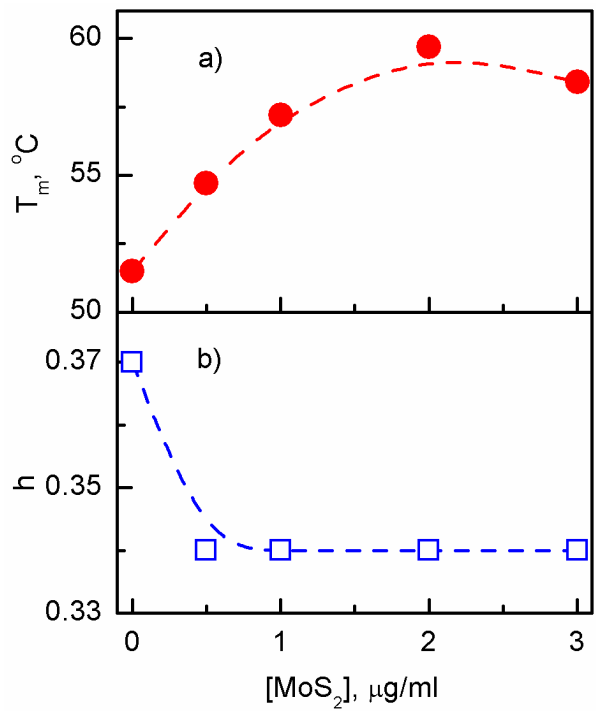

**Fig. 4.** (a) The melting temperature dependence of DNA ($T_m$) as a function of [$c_{MoS2}$]. (b) The concentration dependence of hyperchromic coefficient ($h$) of DNA with MoS$_2$ FLs. The blue and red-dashed lines are B-spline interpolations. The size of the error bar for $T_m$ and $h$ is smaller than that of the symbols.

As most DNA maintains a duplex structure at room temperature in the course of the experiment it means that the nitrogenous bases of double-stranded DNA involved in the system of hydrogen bonds are not available for interaction with MoS$_2$ FLs. This, in turn, makes it difficult for the polymer unwinding. However, looking ahead, our present calculations show that DNA can also bind via covalent binding with the edges of MoS$_2$FL. This effect will be discussed below. The covalent interaction can increase the thermal stability of DNA, which is revealed by the thermal-stability analysis (Fig. 4a).

*3.4. Dynamic light scattering*

In order to obtain more information about the size and morphology of the nanoassemblies, we turn to a DLS analysis for the DNA:MoS$_2$ FL suspension. First, we analyze the MoS$_2$ FL suspension at [$c_{MoS2}$]=1 µg/ml. For such concentration we obtain three different types of the size distribution: the

intensity-, volume-, and number-weighted distributions (from measuring the fluctuation in the scattering intensity), as well as the hydrodynamic diameter distribution (Z-ave or $d_H$), which is derived from the diffusion coefficient (cumulant analysis) (Fig. 5). The TEM analysis of the prepared $MoS_2$ sample reveals that our sample contains mostly FLs (Fig. 1a). It is the most reasonable to use the intensity-based distribution in the characterization of non-spherical species [66]. The intensity-based distribution is also chosen for the analysis of the DNA:$MoS_2$ FL suspension. Figure 6 shows a comparative analysis of the size distributions for the $MoS_2$ FL and DNA:$MoS_2$ FL suspensions taken at $T$=25 °C. A single-size distribution is obtained with a maximum at about 171±4 nm for the $MoS_2$ FL suspension. The estimation of the DLS size of the nanoassemblies formed in the DNA:$MoS_2$ FL suspension gives the average size of about 165±9 nm. The slight difference in the DLS sizes in the $MoS_2$ FLs and the DNA:$MoS_2$ FL nanoassemblies can be explained by the fact that the DNA binding to $MoS_2$ FLs can prevent their stacking.

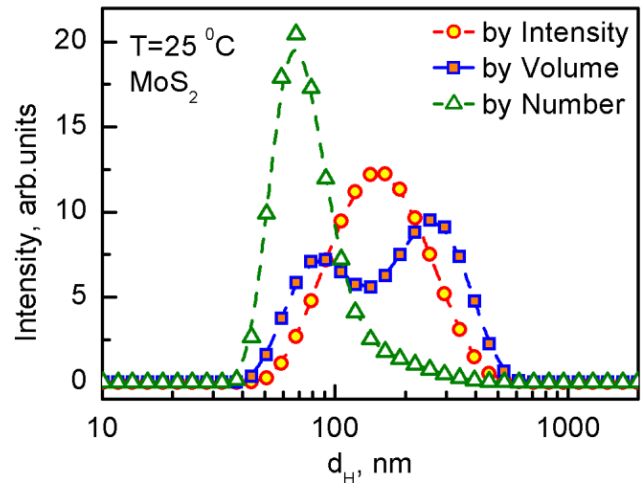

**Fig. 5.** The size distributions by intensity, by volume, and by number for the $MoS_2$ FL colloidal suspension at pH 9.98 at $T$ = 25 °C

The polydispersity index (*PdI*) equal to about 0.21±0.02 suggests that the $MoS_2$ FLs' sample has low polydispersity. The zeta potential of the $MoS_2$ FL suspension is found to be -26±1 mV. The *PdI* value of the DNA:$MoS_2$ FL suspension is slightly larger than for the $MoS_2$ FL suspension and equal to 0.25±0.02. This may be due to the binding of DNA to $MoS_2$ FLs, which results in the blurring of flake outlines. The zeta potential of the nanoassembly suspension is found to be -28±1 mV. The increase in the zeta potential of the DNA:MoS$_2$ FL suspension relative to MoS$_2$ FL one can be explained by the formation of DNA:MoS$_2$ FL nanoassemblies with the total charge being more negative than the total charge of MoS$_2$ FLs since the negatively charged DNA backbone on the of surface MoS$_2$ increased the total charge. In similar experiments, the high ionic strength screens the repulsion of the negatively charged nanostructures facilitating their binding [31]. However, in our experiments the ionic force is rather small. Therefore, another scenario could be involved to explain DNA binding to MoS$_2$ FLs.

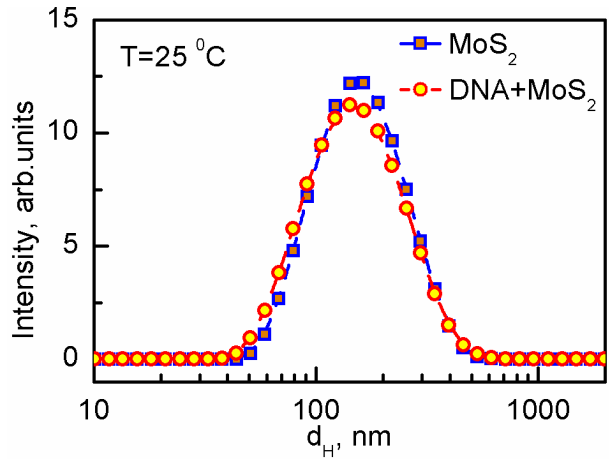

**Fig. 6.** The comparative analysis of the size distributions obtained from the intensity analysis for the MoS$_2$ FL and DNA:MoS$_2$ FL colloidal suspensions taken at $T$=25 °C.

The temperature-dependent DLS measurements of the DNA:MoS$_2$ FL suspension performed in the temperature range of 25–90 °C reveal no size-dependent evolution of the scattering of the nanoassemblies. Figure 7 shows the size-distribution map of the DNA:MoS$_2$ FL nanoassemblies (obtained from the intensity distribution curves) taken in the temperature range from 25 to 90 °C. Upon heating from 25 to 90 °C, the cone saves the bell-shaped profile. The size of the nanoassemblies measured at 90 °C is equal to about 156±9 nm which is slightly less than the estimated size of the nanoassemblies at room temperature. This is explained by a partial decrease in the size of the polymer shell of MoS$_2$ FLs during the DNA unwinding upon heating. The high-temperature size of MoS$_2$ FLs without DNA is closed in the size of the nanoassemblies and equals to 154±6 at 90 °C. The fact that temperature-dependent DLS measurement of the DNA:MoS$_2$ FL suspension did not reveal an aggregation of the nanoassemblies upon heating is not unusual, since DNA:MoS$_2$ FL nanoassemblies are negatively charged and effectively repel each other.

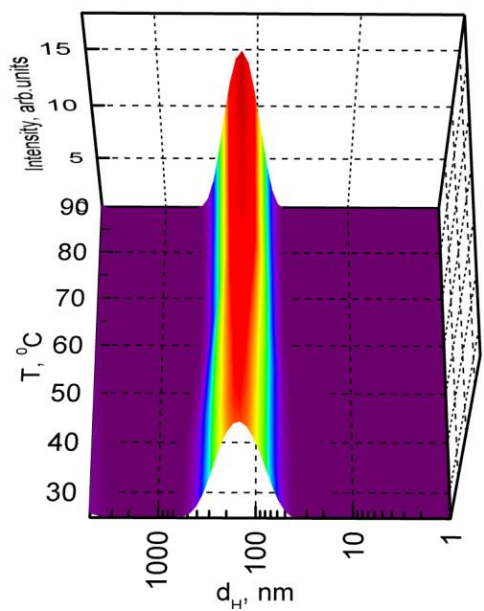

**Fig. 7.** The size-distribution map by intensity for the DNA:MoS$_2$ FL colloidal suspension at pH7 in the temperature range from 25 to 90 °C.

*3.5. Structure and interaction energies of the MoS$_2$-dMP complexes*

The calculated structures of the dMP complexes with the MoS$_2$ nanolayer fragment is shown in Fig. 8. The resulting ZPVE and BSSE corrected interaction energies are collected in Table 1. As it is seen, the interaction energies strongly depend on the structures of complexes. The weakest interaction is observed for the stacked MoS$_2$-dMP S complexes in which the adsorbed molecule is only in contact with the surface of the sulfur atoms. The interaction energies calculated for the stacked complexes are about -23.3 kcal/mol (Table 1). Much stronger interaction is observed for the bonded MoS$_2$-dMP B complexes where the adsorbed molecule interacts with the edge of the MoS$_2$ nanolayer fragment. The calculated interaction energy in the most stable bonded complex is -162.5 kcal/mol. In the bonded complexes, the phosphate group of dMP forms two coordination bonds with the edge molybdenum atoms. Additional hydrogen bonds O-H•••S and C-H•••S are also formed in the complexes. An aqueous environment significantly reduces the interaction energies in all complexes (Table 1). This is especially evident for stacked complexes, for which the interaction energies do not exceed 2 kcal/mol in absolute value. This allows us to conclude that complexes in which the DNA sugar-phosphate backbone interacts with the surface of the MoS$_2$ nanolayer fragments are unstable in aqueous solutions. At the same time, in the covalently bonded complexes, the interaction energy exceeds 50 kcal/mol in absolute value when taking into account the aqueous environment. This indicates that the complexes are stable in water. This is also confirmed by the experiment.

We also investigate the interaction involving dMP with two defects that provide additional possibility for covalent binding with the altered $MoS_2$ surface. In the dMP complex with a disulfur-vacancies $V_{2S}$ defect ($MoS_2(V_{2S})$-dMP B, Fig. 8) there are three coordination bonds formed by oxygen atoms of the phosphate group with the molybdenum atoms. The interaction energy calculated for this complex is -100.7 kcal/mol. Thus, the interaction in this complex is weaker than in the dMP complexes with edge molybdenum atoms. The weakening is caused by steric hindrance of insufficiently large cavity size. It does not allow the phosphate oxygen atoms to approach the molybdenum atoms at the most optimal distance. As the results, the calculated Mo-O bond lengths in the $MoS_2(V_{2S})$-dMP B complex are about 2.4 Å while in the $MoS_2$-dMP B complexes the Mo-O bond lengths are shorter (about 2.1 Å). This results in the energy decrease.

The interaction energy in the dMP complex with the vacancy complex of Mo and its nearest three disulfur pairs ($MoS_2(Mo_S)$-dMP B, Fig. 8) is -101.6 kcal/mol. In this complex, two phosphate oxygen atoms of dMP form coordination bonds with the antisite molybdenum atom. As it is seen, both dMP complexes with the altered $MoS_2$ surface are strongly bonded. Although taking into account the aqueous environment leads to a decrease of the interaction energy to approximately -19.6 kcal/mol in absolute values, these complexes remain stable in water.

In general, the results obtained in the calculations demonstrate that the sugar-phosphate fragment of DNA is capable to form water-stable structures either with the terminal molybdenum atoms or with the molybdenum atoms that become available for interaction as a result of the defects in the $MoS_2$ surface. The theoretical results obtained in this work correlate well with the recent *ab-initio* calculations of the nucleic-acid-bases:$MoS_2$ complexes where it was shown that the interaction energies of the covalently bonded complexes are much higher than the interaction energies of the stacked complexes. Accounting for the effects due to aqueous environment results in a significant decrease in the interaction energies of both types of complexes, but both complexes still remain stable in water [38,39].

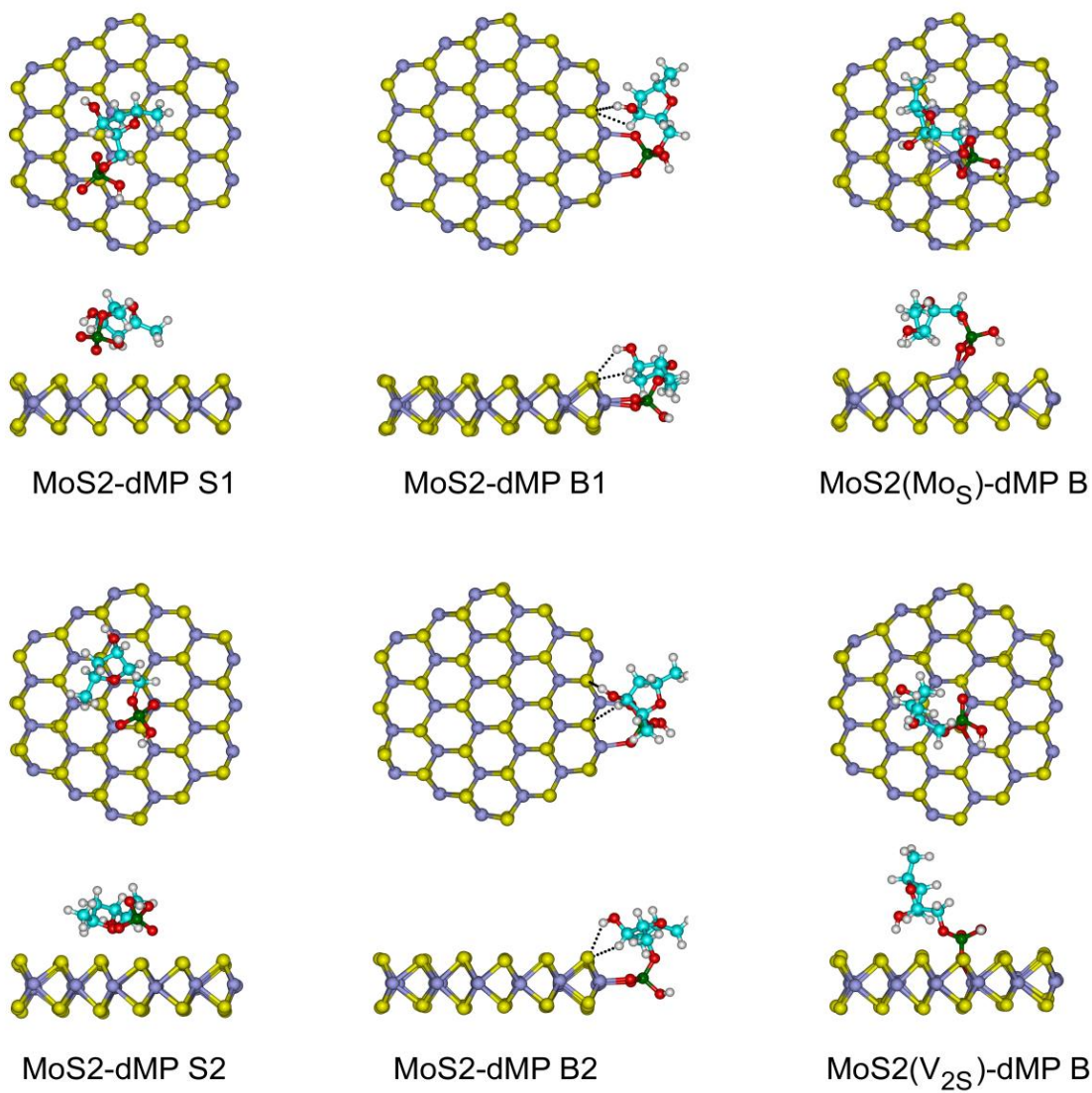

**Fig. 8.** Calculated structure of the MoS$_2$ (pristine layered fragment)–dMP and MoS$_2$ (defects)–dMP complexes (top view and side view): **S** – stacked complexes; **B** – covalently bonded complexes. Complexes with pristine MoS$_2$ are numbered according to their relative stabilities

**Table 1**

Calculated ZPVE and BSSE corrected interaction energies (IE, kcal/mol) of the deoxyribose monophosphate-$MoS_2$ complexes. The interaction energies calculated with accounting for water environment (using the PCM approach) are shown in parentheses.

| Complex | $MoS_2$ | Description[a] | IE |
|---|---|---|---|
| $MoS_2$-dMP S1 | pristine | stacked | -23.3 (-1.2) |
| $MoS_2$-dMP S2 | pristine | stacked | -22.4 (-0.9) |
| $MoS_2$-dMP B1 | pristine | bonded | -162.5 (-54.2) |
| $MoS_2$-dMP B2 | pristine | bonded | -160.9 (-53.7) |
| $MoS_2(Mo_S)$-dMP B | $Mo_S$ | bonded (O-Mo*) | -101.6 (-19.6) |
| $MoS_2(V_{2S})$-dMP B | $V_{2S}$ | bonded (O-Mo**) | -100.7 (-44.3) |

[a], Mo* - molybdenum atom replacing a surface sulfur atom in the $MoS_2$ fragment. Mo** - inner molybdenum atoms available for interaction after the formation of a $V_{2S}$ vacancy.

## 4. Conclusion

A combination of experimental and theoretical methods let us to reveal the binding of native DNA to $MoS_2$ FLs at small ionic strength and pH7. This conclusion is based on the TEM image which showed the adsorption of DNA on $MoS_2$ FLs and the observation of an increase in the thermal stability of DNA and decrease in the hyperchromic coefficient at binding with $MoS_2$ FLs. Observed increase of DNA $T_m$ is explained by the binding of DNA through the phosphate groups with the $MoS_2$ FLs. Consideration of possible complexes of a nucleotide fragment (ribose-phosphate group) with $MoS_2$ nanolayer employing the DFT method showed the forming of the coordination bonds of this nucleotide fragment with Mo atoms located at the edge of the $MoS_2$ nanolayer and with point structural defects of $MoS_2$ surface containing the S vacancy. This observation allows us to propose a mechanism of binding of native DNA to $MoS_2$ FLs, in which their conjugation begins with point contacts of DNA phosphate groups with Mo atoms (at the edge or/and in defects) through the formation of the coordination bond. At this binding, the S-like shape of the DNA melting curve in the nanoassemblies indicates keeping the duplex structure of DNA at the conjugation. Thus, the results obtained indicate the critical role of defects and edge atoms of MoS$_2$FLs in their biofunctionalization and can be used in biosensing and elaboration of new drug delivery systems.

**Author contributions**

All authors discussed the results and commented on the manuscript. I.V. and V.V. prepared the samples. V.V. and E.U. carried out the spectroscopic measurements, thermal denaturation, and analyzed data. A.S. and A.L. performed and analyzed the DLS data. S.P. performed TEM characterization. S.S. and L.A. provided the theoretical study. A.G., S.S and V.K. planned and coordinated the project.

**Acknowledgements**

The authors acknowledge partial financial support from National Academy of Sciences of Ukraine (Grant №0123U100628). A.L. expresses gratitude to the Ministry of Education and Science of Ukraine for partial financial support in the frame of Project №0124U000968. E.U. acknowledges Wolfgang Pauli Institute, Vienna, Austria for the financial support (Pauli Postdoc research training scholarship in the field of Data analysis in molecular biophysics in the context of the WPI thematic program "Numerical models in Biology and Medicine" (2023/2024). A.G. acknowledges Wolfgang Pauli Institute, Vienna, Austria for the financial support (Pauli Postdoc research training scholarship in the field of Data analysis in molecular biophysics in the context of the WPI thematic program "Mathematics for Biology and Medicine" (2024/2025). E.U. and A.G. acknowledge Nanophotonics journal, De Gruyter, Sciencewise Publishing, and the Optica Foundation for the financial support (Ukraine Optics and Photonics Researcher Grant 2023). An allocation of computer time from the Computational Center at Institute for Low Temperature Physics and Engineering and from UA Research High Performance Computing (HPC) and High Throughput Computing (HTC) at the University of Arizona is gratefully acknowledged.

**Supporting Information**

The Supporting Information to this article can be found online at